\documentclass{llncs}
\usepackage{xspace}
\usepackage{color}
\usepackage{todonotes}
\usepackage{subfigure}
\usepackage{hyperref}
\usepackage[capitalise]{cleveref}
\usepackage{marvosym}
\newcommand{\orcid}[1]{\href{https://orcid.org/#1}{\includegraphics[scale=1]{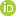}}}

\let\doendproof\endproof
\renewcommand\endproof{~\hfill$\qed$\doendproof}

\newcommand{\sys}{\textsc{StoryTreeViewer}\xspace}

\newcommand{\sua}{SUA\xspace}
\renewcommand{\S}{\mathcal{S}}

\newcommand{\G}{\mathcal{G}}
\newcommand{\A}{\mathcal{A}}

\begin{document}

\title{Storyline Visualizations with Ubiquitous Actors\thanks{This work is partially supported by: $(i)$ MIUR, grant 20174LF3T8 ``AHeAD: efficient Algorithms for HArnessing networked Data'', $(ii)$ Dipartimento di Ingegneria - Universit\`a degli Studi di Perugia, grant RICBA19FM: ``Modelli, algoritmi e sistemi per la visualizzazione di grafi e reti''.}}
\author{Emilio Di Giacomo\texorpdfstring{ \href{https://orcid.org/0000-0002-9794-1928}{\protect\includegraphics[scale=0.45]{orcid}}}{} \and Walter Didimo\texorpdfstring{ \href{https://orcid.org/0000-0002-4379-6059}{\protect\includegraphics[scale=0.45]{orcid}}}{} \and Giuseppe Liotta\texorpdfstring{ \href{https://orcid.org/0000-0002-2886-9694}{\protect\includegraphics[scale=0.45]{orcid}}}{} \and \\Fabrizio Montecchiani\texorpdfstring{ \href{https://orcid.org/0000-0002-0543-8912}{\protect\includegraphics[scale=0.45]{orcid}}}{} \and Alessandra Tappini\texorpdfstring{ \href{https://orcid.org/0000-0001-9192-2067}{\protect\includegraphics[scale=0.45]{orcid}}}{}\textsuperscript{(\Letter)}}


\institute{Engineering Department, University of Perugia, Perugia, Italy\\\email{name.surname@unipg.it}
}
\maketitle
\begin{abstract}
    Storyline visualizations depict the temporal dynamics of social interactions, as they describe how groups of actors (individuals or organizations) change over time. A common constraint in storyline visualizations is that an actor cannot belong to two different groups at the same time instant. However, this constraint may be too severe in some application scenarios, thus we generalize the model by allowing an actor to simultaneously belong to distinct groups at any point in time. We call this model \emph{Storyline with Ubiquitous Actors} ({\em \sua}). Essential to our model is that an actor is represented as a tree rather than a single line. We describe an algorithmic pipeline to compute storyline visualizations in the \sua model and discuss case studies on publication data.
\end{abstract}


\section{Introduction}\label{se:intro}
Storyline visualizations have been the focus of intense research in the last decade. Originally introduced to describe the narrative of a movie~\cite{Munroe}, this visualization paradigm has been successfully used to represent the temporal dynamics of the interactions between actors (individuals or organizations) in a social network or in a working environment (\cite{DBLP:journals/tvcg/LiuWWLL13,DBLP:conf/softvis/OgawaM10,DBLP:journals/cg/PadiaBH19,DBLP:conf/cw/QiangB16,Tanahashi2012,DBLP:journals/tvcg/TanahashiHM15,DBLP:journals/information/TongRBWLWLWQLM18}). 
%
%
In a storyline visualization, the narrative unfolds from left to right, each actor is represented as a line, and two lines may converge or diverge at a time instant based on whether the two corresponding actors interact or not at that instant; see~\cref{fi:storyline-classic}. Since a group of lines bundled together usually reflects an in-person meeting, a common constraint in a storyline visualization is that an actor cannot belong to two different groups at the same point in time. However, this constraint represents a severe limitation for some application scenarios, for example when groups model associations that are not in-person meetings (e.g., paper co-authorships) or when each point in time of the storyline corresponds to a relatively long time interval~(e.g., one~year). 

In this paper we generalize the classical storyline model by allowing an actor to simultaneously belong to distinct groups. We call this model \emph{Storyline with Ubiquitous Actors}~({\em \sua}); see~\cref{fi:storyline-tree}. Essential to our model is that an actor is represented as a tree rather than a single line. Our contribution is: $(i)$ We propose a visualization paradigm for the \sua model and identify quality metrics for it. $(ii)$ We define an algorithmic pipeline for storyline visualizations in the \sua model. $(iii)$ We provide a proof-of-concept implementation and apply it to produce visualizations in real-life scenarios.

\medskip\noindent{\bf Related Work.} Tanahashi and Ma~\cite{Tanahashi2012} present a general framework for generating aesthetically pleasing storyline visualizations. Subsequent papers focus on specific optimization problems like crossing minimization~\cite{vanDijk2018,DBLP:conf/gd/GronemannJLM16,DBLP:conf/gd/KostitsynaNP0S15} and wiggle minimization~\cite{fn-mwsv-2017}.
Padia et al.~\cite{DBLP:conf/graphicsinterface/PadiaBH18,DBLP:journals/cg/PadiaBH19} consider storyline visualizations with multiple timelines. Efficient approaches that compute storyline visualizations with hierarchical relationships or with streaming data are described by Liu et al~\cite{DBLP:journals/tvcg/LiuWWLL13} and by Tanahashi et al.~\cite{DBLP:journals/tvcg/TanahashiHM15}, respectively. Qiang and Bingjie~\cite{DBLP:conf/cw/QiangB16} present a system that embeds storyline visualizations into a radial layout.
For a broader dissertation on storytelling~and~visualization refer to the survey of Tong et al.~\cite{DBLP:journals/information/TongRBWLWLWQLM18}.
We~remark that our scenario is strongly related to the dynamic sets visualization; see, e.g.,~\cite{DBLP:journals/cgf/MizunoWTI19,DBLP:journals/ivs/NguyenXWW16}, and~\cite{DBLP:journals/cgf/AlsallakhMAHMR16} for a survey. In this regard, it is worth mentioning a recent work by Agarwal and Beck~\cite{doi:10.1111/cgf.13988}, who adopt storylines for visualizing~dynamic~sets.


\section{Storyline Visualizations and Ubiquitous Actors}\label{se:model}
We first recall basic definitions and principles of classical storyline visualizations and then define our visualization for the \sua model. 

\smallskip\noindent\textbf{Classical storyline visualizations.} A \emph{storyline} $\S = (\A,\G)$ consists of a set $\A=\{a_1, a_2, \dots, a_n\}$ of \emph{actors} and a set $\G = \{G_1, G_2, \dots, G_k\}$ of \emph{groups}. Each group $G_i \in \G$ is a triple $\langle\A(G_i), b_i, e_i\rangle$, where $\A(G_i) \subseteq \A$ is a subset of actors, $b_i$ is the \emph{begin-time} of $G_i$ and $e_i$ is the \emph{end-time} of $G_i$. We say that $G_i$ is \emph{active} at any time instant in the interval $[b_i,e_i]$, and that each actor $a_j \in \A(G_i)$ \emph{participates} to $G_i$.
A common assumption is that an actor cannot participate to two distinct groups at the same point in time, i.e., if $\G_i$ and $\G_j$ are two distinct groups such that $[b_i,e_i] \cap [b_j,e_j] \neq \emptyset$ then $\A(G_i) \cap \A(G_j) = \emptyset$.

\begin{figure}[tb]
    \centering
    \subfigure[]{\label{fi:storyline-classic}\includegraphics[width=0.48\columnwidth,page=1]{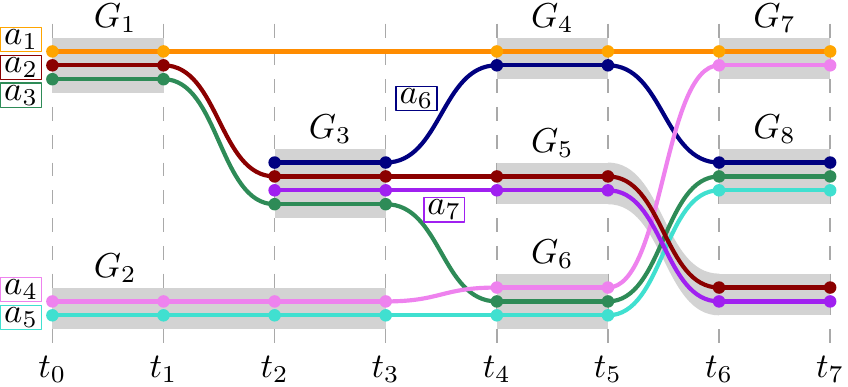}}
    \subfigure[]{\label{fi:storyline-tree}\includegraphics[width=0.48\columnwidth,page=2]{storyline}}
    \caption{Storyline visualization: (a) Classical model. (b) \sua model.}
    \label{fi:storyline}
\end{figure}

In a \emph{storyline visualization}, each actor $a_j$ is represented as a line $\ell_j$ that flows from left to right; see~\cref{fi:storyline-classic}. Some basic \emph{principles} are considered: $(i)$ For each group $G_i$, the lines representing the actors in $\A(G_i)$ are \emph{adjacent}, i.e., they run close together from the begin-time $b_i$ to the end-time $e_i$ of $G_i$; $(ii)$ lines of actors that are not in the same group at the same time are depicted relatively far from one another; $(iii)$ a line should not deviate unless it converges or diverges with another line.
In addition, common \emph{quality metrics} for the readability of storyline visualizations are: (a) \emph{Line or block crossings} -- a line crossing occurs when two lines intersect while a block crossing is caused by two blocks of parallel lines that pairwise intersect. (b) \emph{Line wiggles} -- line deviations that, when frequent, negatively affect the visual flow of the layout.
(c) \emph{White space gaps} -- white areas used to separate lines of actors that do not participate to the same group.



\smallskip\noindent\textbf{Visualizations with ubiquitous actors.} 
To support the visualization of ubiquitous actors, we represent an actor $a_j$ as a tree $\tau_j$ rather than as a line (see~\cref{se:algorithms} for a formal definition of $\tau_j$). Informally speaking, when an actor simultaneously participates to different groups, the line of the actor branches out and forms a tree. For example, in \cref{fi:storyline-tree} we see the trees of five actors $a_1, \dots, a_5$. At time $t_1$ actor $a_2$ participates to group $G_1$ while at time $t_2$ it simultaneously participates to groups $G_1$ and $G_3$. As a consequence, the line of $a_2$ at time $t_1$ is split into two branches.
The choice of a tree is motivated by the fact that we want to represent each actor by a connected geometric feature (avoiding discontinuities); at the same time, we want to keep such geometric feature as simple as possible, since the addition of edges may increase the number of crossings.
%
%
Such tree representations add new quality metrics: 

        \noindent-- {\em Actor planarity.} It is natural to require that each tree representing an actor is not self-intersecting. While this is trivially guaranteed when an actor is a line, it requires an algorithmic effort in the \sua model.

        \noindent-- {\em Branch continuity.} To avoid interruptions in the continuity of the story, the number of branches of an actor at time $t_h$ that continue at time $t_{h+1}$ should be maximized. If an actor participates to $m$ groups at time $t_h$ and to $m' \geq m$ groups at time $t_{h+1}$, all branches at time $t_h$ should continue at time $t_{h+1}$.
       
        \noindent-- {\em Branch degree.} When an actor tree needs new branches at some time instant~$t_h$, it is desirable that the maximum number of branches that emanates from a common branch at time~$t_{h-1}$ is minimized. 

\begin{figure}[t]
    \centering
    \subfigure[]{\label{fi:conflict-1.a}\includegraphics[width=0.48\columnwidth,page=5]{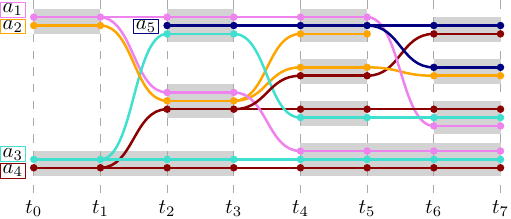}}\hfill
    \subfigure[]{\label{fi:conflict-1.b}\includegraphics[width=0.48\columnwidth,page=6]{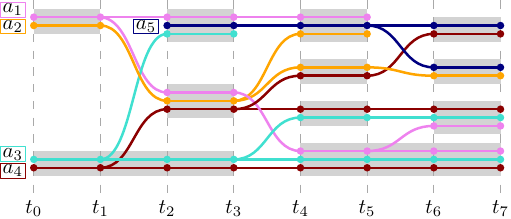}}
    \subfigure[]{\label{fi:conflict-2.a}\includegraphics[width=0.48\columnwidth,page=7]{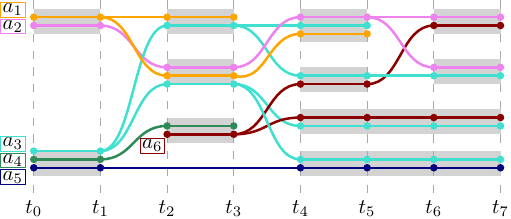}}\hfill
    \subfigure[]{\label{fi:conflict-2.b}\includegraphics[width=0.48\columnwidth,page=8]{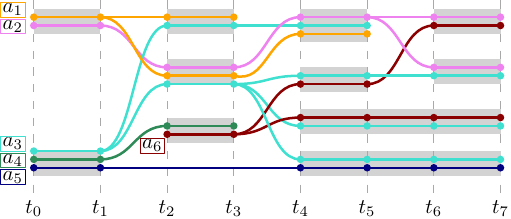}}
    \caption{(a) A layout with optimal branch continuity. (b) Violating branch continuity for $a_3$ at $t_3$ and for $a_1$ at $t_5$ removes $9$ crossings and reduces line wiggles. (c) A layout with optimal branch degree. (d) Violating branch degree for $a_3$ at $t_3$ reduces line wiggles and removes $2$ crossings.}
    \label{fi:conflict-2}
\end{figure}

\noindent We note that such new metrics may be in conflict with classical ones (see~\cref{fi:conflict-2}). 


\section{The \sua Algorithmic Pipeline}\label{se:algorithms}

We compute storyline visualizations in the \sua model by means of an algorithmic  pipeline based on the  concept of \emph{actor-tree} $\tau_j$ associated with an actor $a_j$. The \emph{life-time} of actor $a_j$ is the interval between the first and the last time instant at which $a_j$ belongs to some group. Tree  $\tau_j$ is defined as follows.
%
%
 %
\textsf{Node set --} Tree $\tau_j$ has a root $r_j$. For each time instant $t_h$ in the life-time of $a_j$: If $a_j$ participates to at least one group $G_i$ ($i>0$) active at $t_h$, $\tau_j$ has a node $u_{h,i}$ for each such group; otherwise $\tau_j$ has a single node $u_{h,0}$ that is not associated with any group.  
\textsf{Edge set --} The parent of a node $u_{h,i}$ ($i \geq 0$) is assigned as follows: If $t_{h-1}$ is not in the life-time of $a_j$, the parent of $u_{h,i}$ is the root $r_j$. Else, if $i>0$ and $G_i$ is active before time $t_h$, the parent of $u_{h,i}$ is $u_{h-1,i}$. Else, the parent of $u_{h,i}$ is one of the nodes $u_{h-1,l}$.
%
Our algorithmic pipeline consists of four steps:

\smallskip\noindent{\sf 1. Actor-tree Initialization}. It defines an initial actor-tree $\tau_j$ for each actor $a_j$. Namely, given the nodes of $\tau_j$, it assigns the parent to each node of $\tau_j$.

\smallskip\noindent{\sf 2. Branch Permutation}. For each time instant $t_h$ this step computes a permutation (i.e., a vertical order) of all nodes at time $t_h$ in the union of all actor-trees .

\smallskip\noindent{\sf 3. Actor-tree Untangling}. For each actor-tree $\tau_j$, it redefines the parent of some nodes, so to reduce self-intersections of $\tau_j$ without changing its node degrees.   

\smallskip\noindent{\sf 4. Branch-coordinate Assignment}. It assigns the $y$-coordinates to actor-tree nodes.

\begin{figure}[ht]
    \centering
    \subfigure[\scriptsize \textsf{Actor-tree Initialization}: Input]{\includegraphics[width=0.45\columnwidth, page=13]{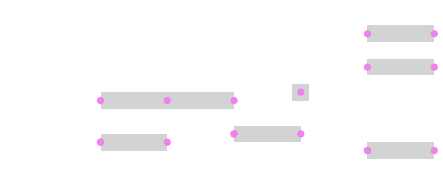}\label{fi:alg-pipeline-1}}\hfil
    \subfigure[\scriptsize \textsf{Actor-tree Initialization}: Output]{\includegraphics[width=0.45\columnwidth, page=12]{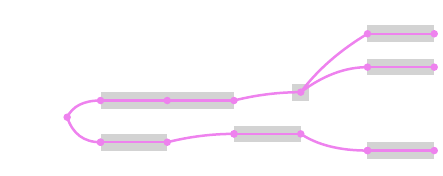}\label{fi:alg-pipeline-2}}
     \subfigure[\scriptsize \textsf{Branch Permutation}: Disjoint paths]{\includegraphics[width=0.45\columnwidth, page=11]{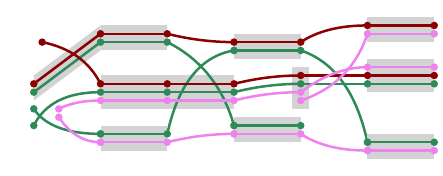}\label{fi:alg-pipeline-5}}\hfil
    \subfigure[\scriptsize \textsf{Branch Permutation}: Algorithm in~\cite{vanDijk2018}]{\includegraphics[width=0.48\columnwidth, page=6]{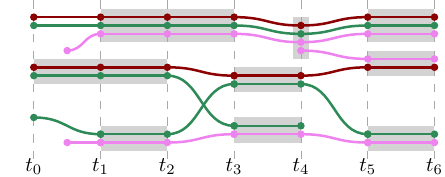}\label{fi:alg-pipeline-6}}
    \subfigure[\scriptsize \textsf{Branch Permutation}: Output]{\includegraphics[width=0.45\columnwidth, page=7]{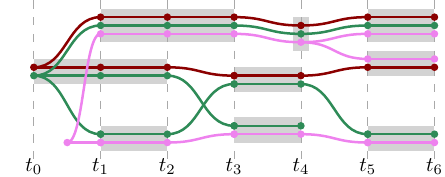}\label{fi:alg-pipeline-7}}\hfil
    \subfigure[\scriptsize \textsf{Actor-tree Untang.} +  \textsf{Branch-coord. Assign.}]{\includegraphics[width=0.45\columnwidth, page=8]{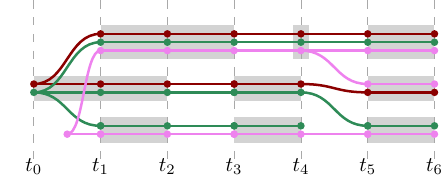}\label{fi:alg-pipeline-8}}
    \caption{ Illustration of the algorithmic pipeline.
    } \label{fi:alg-pipeline}
\end{figure}

In Step~1 we aim to optimize branch continuity and branch degree. Step~2 aims to minimize block or line crossings. Step~3 tries to enforce actor planarity. Step~4 aims to reduce line wiggles and space gaps.
Different algorithmic strategies are applicable to each step. We briefly describe our solution; see also \cref{fi:alg-pipeline}.

\smallskip\noindent {\sf Actor-tree Initialization.} For any actor-tree $\tau_j$, let $V_{h-1}$ and $V_h$ be the sets of nodes at time instant $t_{h-1}$ and $t_h$. The parents of the nodes in $V_h$ are chosen among the nodes in $V_{h-1}$ so that the distribution of the degrees in $V_{h-1}$ is as uniform as possible. For each node $u_{h,i}$ in $V_h$ that belongs to the same group of a node $u_{h-1,i}$ in $V_{h-1}$, the parent of $u_{h,i}$ is $u_{h-1,i}$. For the remaining nodes of $V_h$, we adopt a round robin policy to assign children to the nodes in $V_{h-1}$ so that the difference of the degrees of any two nodes of $V_{h-1}$ is~at~most~one. 

\begin{figure}[t]
    \centering
    \subfigure[]{\label{fi:algo-steps-a}\includegraphics[width=0.48\columnwidth,page=1]{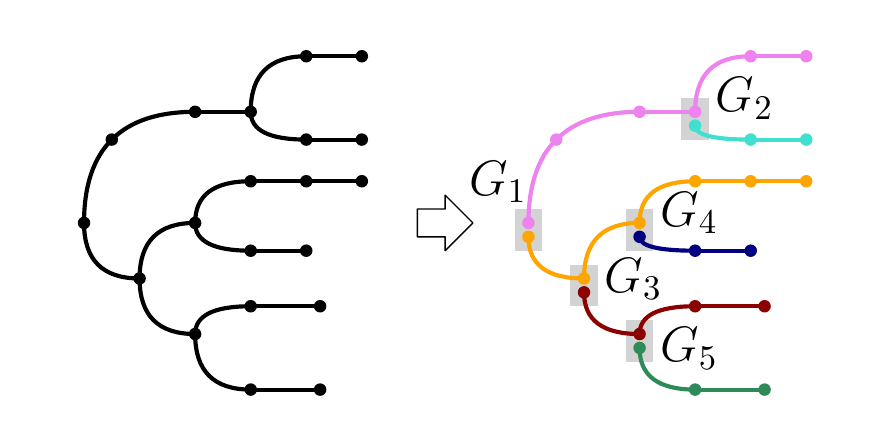}}\hfill
    \subfigure[]{\label{fi:algo-steps-b}\includegraphics[width=0.12\columnwidth,page=2]{algo-steps}}\hfill
    \subfigure[]{\label{fi:algo-steps-c}\includegraphics[width=0.24\columnwidth,page=3]{algo-steps}}\hfill
    \subfigure[]{\label{fi:algo-steps-d}\includegraphics[width=0.12\columnwidth,page=4]{algo-steps}}
    \caption{(a) Transformation of an actor tree into a set of disjoint paths. (b) Crossing removal when merging two copies of the same node. (c) Actor-tree untangling. (d) Preservation of the edge order around a node when splitting it. }
    \label{fi:algo-steps}
\end{figure}

\smallskip\noindent \textsf{Branch Permutation.} We exploit a state-of-the-art algorithm for classical storyline visualizations, namely the algorithm by van Dijk et al.~\cite{vanDijk2018} based on a SAT formulation, which optimally solves the problem of minimizing block crossings. To this aim, we transform the output of Step~1 into an instance for a classical storyline visualization: Each actor tree is partitioned into a set of edge-disjoint paths by duplicating each node with $k \geq 2$ children into $k$ nodes each having one child (see~\cref{fi:algo-steps-a}). Each path is processed by the algorithm in~\cite{vanDijk2018} as a distinct actor.
All copies of the same node are then recombined into a single node to restore the tree. However, if disjoint paths originating from two copies of the same node are treated independently, they can create many crossings when recombined back into the tree.
To alleviate this drawback, we let the initial node of each path belong to the same group of its original duplicate, unless this operation makes the path belonging to multiple groups at some other point in time. Moreover, when copies of the same node are recombined into a single node, two edges incident to this node may create a crossing, which is easily removed as depicted in~\cref{fi:algo-steps-b}.
Hence, a crossing in an actor tree~only~involves~independent~edges.


\smallskip\noindent \textsf{Actor-tree Untangling.} For each actor-tree $\tau_j$, we redefine the parent of some nodes to reduce the number of crossings between the edges of $\tau_j$. If two edges $(u_{h-1,p},u_{h,q})$ and $(u_{h-1,r},u_{h,s})$ ($p \neq q$, $r \neq s$) of $\tau_j$ cross, we replace them with two new edges $(u_{h-1,p},u_{h,s})$ and $(u_{h-1,r},u_{h,q})$ (see~\cref{fi:algo-steps-c}). This operation removes at least one self-intersection and does not create any new one. Also, the degree of $u_{h-1,p}$, $u_{h,q}$, $u_{h-1,r}$, and $u_{h,s}$ does not change. We repeat this procedure until it is no longer possible to remove self-intersections from $\tau_j$. 

\smallskip \noindent\textsf{Branch-coordinate Assignment.} As in the \textsf{Branch-permutation} step, we consider the set of paths that decompose the tree and make them disjoint by duplicating each node with $k \geq 2$ children into $k$ nodes each having one child. The cyclic order of the edges around each node defines the permutation of the lines that correspond to these edges (see \cref{fi:algo-steps-d}). Any technique that assigns coordinates to the paths, while reducing line wiggles and white space gaps can be applied (see, e.g.,~\cite{fn-mwsv-2017,Tanahashi2012}). This assignment preserves the vertical permutations of the~paths.

\section{Implementation and Case Studies}\label{se:experiments}

We developed a prototype web application, \sys, which implements the algorithmic pipeline of \cref{se:algorithms}, see \texttt{\small{\url{https://bit.ly/2yS3Fvi}}}.     
\sys offers a simple interactive interface, which we used to evaluate effectiveness and limits of our model through two case studies on publication data extracted from DBLP~\cite{dblp} and Scopus~\cite{scopus}.


    
    
    

\begin{figure}[t]
    \centering
    \subfigure[Case Study 1]{\includegraphics[width=\columnwidth]{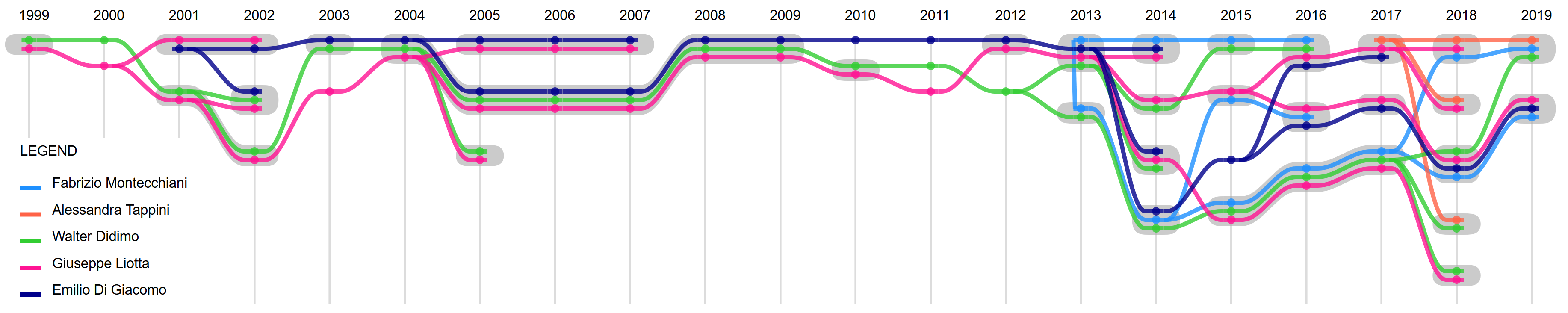}\label{fi:case-study-1}}\hfil
    \subfigure[Case Study 2]{\includegraphics[width=\columnwidth]{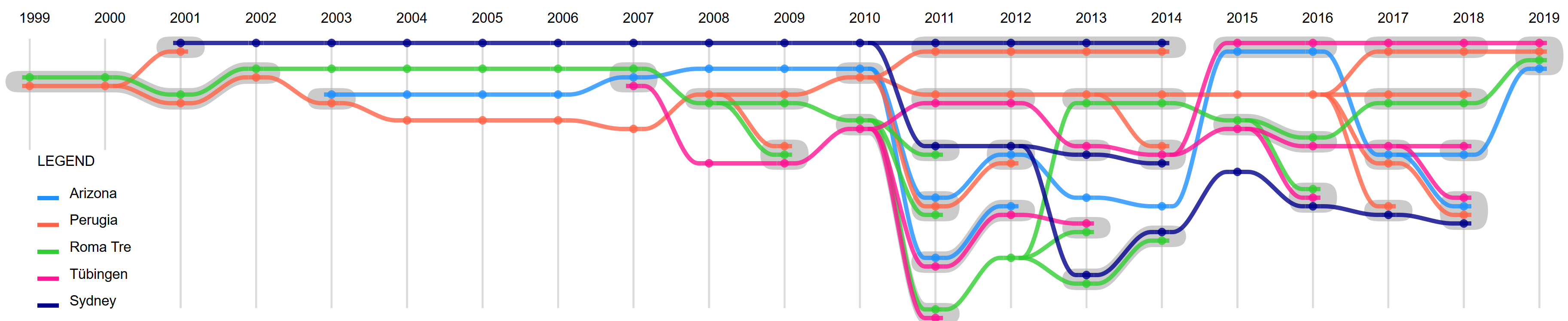}\label{fi:case-study-2}}
    \caption{Visualizations of our case studies. See the appendix for larger images.}
\end{figure}

\smallskip\noindent\textbf{Case study 1.} 
The first case study, see \cref{fi:case-study-1}, describes scientific collaborations among the authors of this work in the various editions of the Graph Drawing Symposium (GD) since 1999. Each actor is an author and a group $G_i$ is a subset of actors who co-authored some papers. $G_i$ is active in $[b_i,e_i]$ if all their members co-authored at least one paper in each year from $b_i$ to $e_i$.
The layout reveals the following dynamic. In the first part of the story there is a strong collaboration between the three oldest actors (pink, green, and blue), in particular they form a group lasting from 2004 to 2009. In 2003, the pink actor was the chair of GD, which prevented him to publish together with the other two authors. In 2010 and 2011 the collaboration of the three actors is weaker, as they mainly collaborated with researchers outside their university.
%
%
The dynamic becomes more involved in the last years, when two new members joined~the~group (cyan and orange), and new theoretical and application research topics were~activated.

\smallskip\noindent\textbf{Case study 2.}
The second case study, see \cref{fi:case-study-2}, describes scientific collaborations among five of the research teams (universities) with the highest number of papers published at GD. The actors are the teams and the groups are defined as in case study 1. Namely, a group $G_i$ is a subset of teams that appear together in some papers (in terms of author affiliations); $G_i$ is active in $[b_i,e_i]$ if all its teams appear together in at least one paper in each year from~$b_i$~to~$e_i$.
The layout shows some interesting facts. From 1999 to 2002 there is a strong collaboration between Roma Tre and Perugia, witnessing that the group in Perugia stems from researchers coming from Rome. The collaboration between the five research teams increases since 2007 and becomes stronger since 2011. This is partly explained by the series of workshops  started around 2006 (BWGD, HOMONOLO, GNV, etc.) that increased international collaborations.

%

\smallskip\noindent\textbf{Limits.} Working on the case studies, we observed some limits of our approach: $(i)$ The implementation for the \textsf{Branch Permutation} step exploits the algorithm in~\cite{vanDijk2018}, splitting each actor-tree into multiple disjoint paths. The size of this transformed instance raises some computational complexity issues. $(ii)$ Our visualizations appear to be readable for relatively few actors and further work is needed to better evaluate the effectiveness of the \sua model on larger instances.


\section{Conclusions and Future Work}\label{se:conclusions}
We introduced the \sua model, which allows ubiquitous actors in storyline visualizations. This model extends the spectrum of applications for this type of representation and opens up to many intriguing research directions. Among them: $(i)$ Are there more effective ways of modeling ubiquitous actors other than using trees? $(ii)$ Design and experiment different algorithms for the \sua pipeline.

\newpage
\bibliography{biblio}
\bibliographystyle{splncs04}

\appendix

\section*{Appendix}\label{ap:appendix}

\begin{figure}[h]
    \centering
    \subfigure[Case study 1]{\includegraphics[height=0.3\columnwidth,angle=-90]{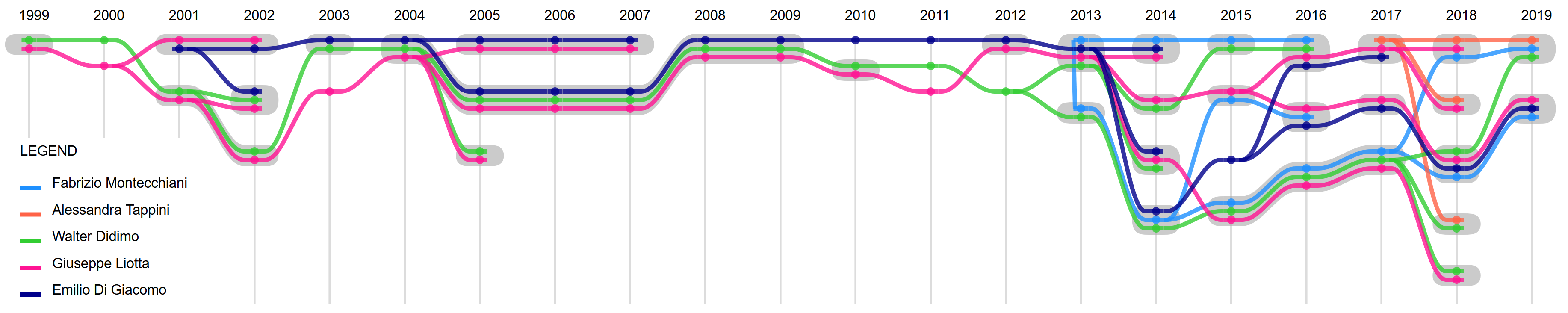}\label{fi:case-study-1-big}}\hfil
    \subfigure[Case study 2]{\includegraphics[height=0.307\columnwidth,angle=-90]{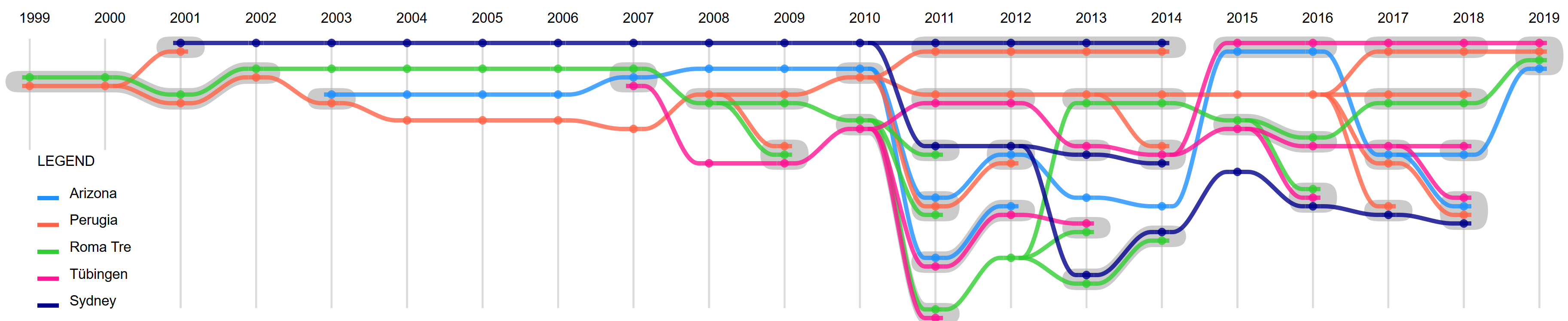}\label{fi:case-study-2-big}}
    \caption{Visualizations of our case studies.}
\end{figure}

\end{document}